\documentclass{article}

\usepackage{arxiv}
\usepackage{array}

\usepackage[utf8]{inputenc} 
\usepackage[T1]{fontenc}    
\usepackage{hyperref}       
\usepackage{url}            
\usepackage{booktabs}       
\usepackage{graphicx}

\title{Early Warning of COVID-19 Hotspots using Mobility of High Risk Users from Web Search Queries}

\author{
 Takahiro Yabe \\
  Lyles School of Civil Engineering\\
  Purdue University\\
  \texttt{tyabe@purdue.edu} \\
   \And
 Kota Tsubouchi \\
  Yahoo Japan Corporation\\
  Tokyo, Japan\\
  \texttt{ktsubouc@yahoo-corp.jp} \\
  \And
 Satish V Ukkusuri \\
  Lyles School of Civil Engineering\\
  Purdue University\\
  \texttt{sukkusur@purdue.edu} \\
}

\begin{document}
\maketitle
\begin{abstract}
COVID-19 has disrupted the global economy and well-being of people at an unprecedented scale and magnitude. 
To contain the disease, an effective early warning system that predicts the locations of outbreaks is of crucial importance. 
Studies have shown the effectiveness of using large-scale mobility data to monitor the impacts of non-pharmaceutical interventions (e.g., lockdowns) through population density analysis. 
However, predicting the locations of potential outbreak occurrence is difficult using mobility data alone. 
Meanwhile, web search queries have been shown to be good predictors of the disease spread. 
In this study, we utilize a unique dataset of human mobility trajectories (GPS traces) and web search queries with common user identifiers (> 450K users), to predict COVID-19 hotspot locations beforehand. 
More specifically, web search query analysis is conducted to identify users with high risk of COVID-19 contraction, and social contact analysis was further performed on the mobility patterns of these users to quantify the risk of an outbreak.
Our approach is empirically tested using data collected from users in Tokyo, Japan. 
We show that by integrating COVID-19 related web search query analytics with social contact networks, we are able to predict COVID-19 hotspot locations 1-2 weeks beforehand, compared to just using social contact indexes or web search data analysis. 
This study proposes a novel method that can be used in early warning systems for disease outbreak hotspots, which can assist government agencies to prepare effective strategies to prevent further disease spread.
\end{abstract}


\section{Introduction}
The coronavirus pandemic (COVID-19) has inflicted significant health and economic impacts across the globe. 
Due to the contagious nature of the disease, monitoring and controlling how people come in contact with each other has shown to be key in containing the disease \cite{zhang2020changes}. 
Therefore, in response to the COVID-19 pandemic, countries have taken various non-pharmaceutical interventions (NPIs) (e.g., lockdowns, social distancing, testing and tracing) \cite{flaxman2020report}, while facing the conundrum of balancing the trade-off between benefits on public health and depletion of economic performance \cite{satishpolicy}. 
In order to monitor and evaluate the impacts of such interventions, large scale mobile phone data has been utilized as an effective data source \cite{oliver2020mobile}. 
Studies on human mobility analysis have used such data to model disease dynamics \cite{bengtsson2015using,finger2016mobile,tizzoni2014use,wesolowski2012quantifying}.
During the COVID-19 crisis, various stakeholders have utilized large-scale mobility datasets to evaluate the effects of NPIs in various regions \cite{klein2020assessing,kraemer2020effect,lai2020effect,pepe2020covid,bonato2020mobile,wellenius2020impacts,gao2020mapping,dahlberg2020effects,santana2020analysis,cintia2020relationship,schlosser2020covid19,yabe2020non}.

\begin{figure}[t]
\centering
\includegraphics[width=.65\columnwidth]{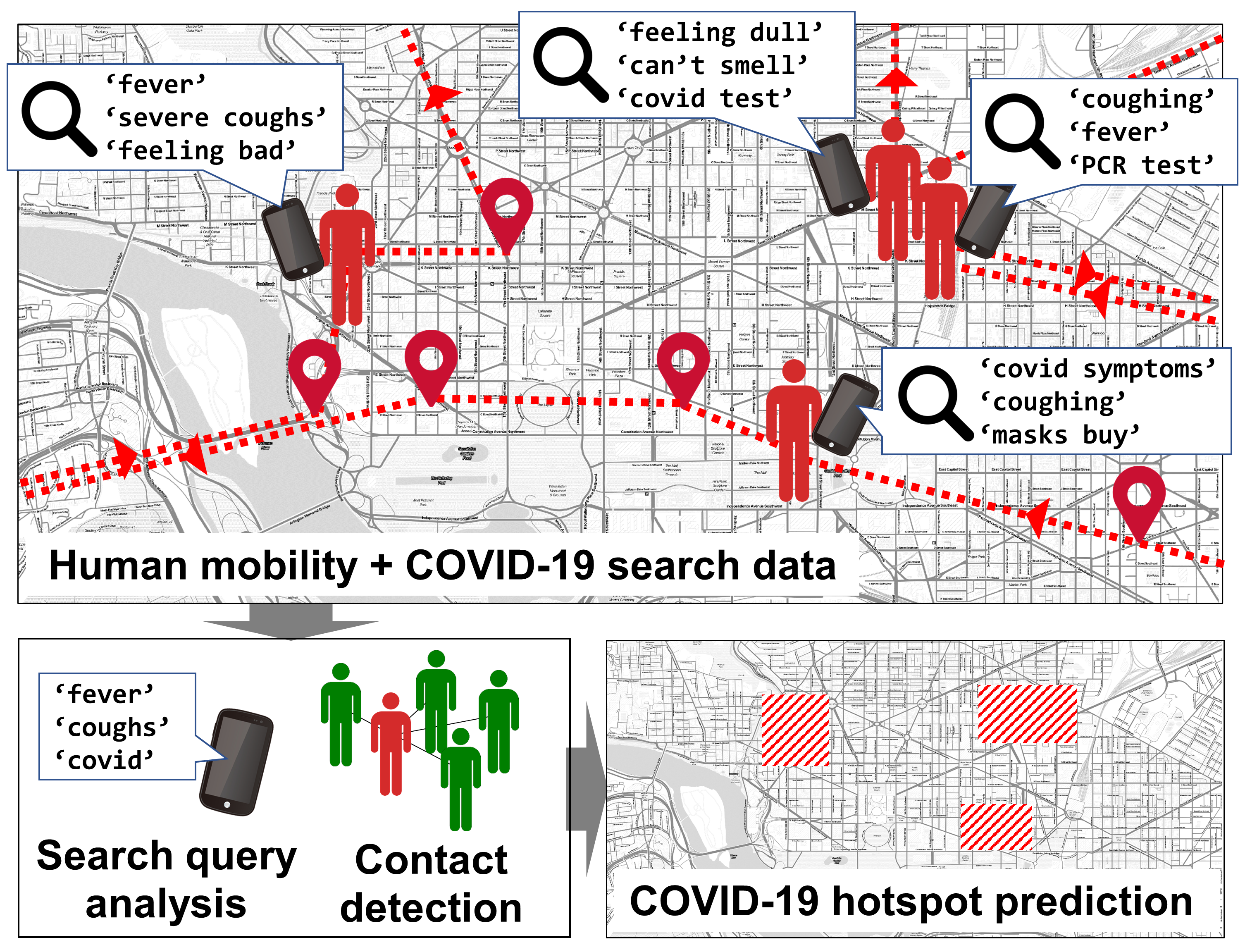}
\caption{Overview of the study. Using individual human mobility trajectories and web search data with common user IDs, we aim to predict COVID-19 hotspot locations via search query analysis and social contact detection.} 
\label{overview}
\end{figure}

The aforementioned studies have shown the effectiveness of using large-scale mobility data to monitor physical co-location of the population (which can be used as a proxy for social contacts) in a fine-grained spatial and temporal scale.
More specifically, metrics such as ``social contact index (SCI)'' have been proposed, and their strong relationships with the estimated transmissibility of COVID-19 (i.e. effective reproduction number $R_t$) has been shown \cite{yabe2020non}. 
However, such metrics are incapable of predicting outbreaks beforehand, since such analysis can only be conducted in a retrospective manner. 
This is a critical drawback since countries are already facing second and third waves of COVID-19 (as of October 2020), and countries are in need of effective early warning systems that can predict when and where the next outbreaks would occur beforehand. 

To tackle this problem, we utilize a unique dataset that contains both the GPS location data (mobility trajectories) and web search queries of users, which are linked with common user identifiers (IDs). 
We hypothesize that users who search COVID-19 symptom related queries more frequently and intensely have a higher risk of having contracted the virus. 
By integrating the mobility analysis and web search analysis, we propose the ``high-risk social contact index (HR-SCI)'' metric, which takes into consideration both the density of population and risk of COVID-19 (Figure \ref{overview}), in contrast to previous metrics that only rely on either mobility data or web search data alone. 
The methods are tested using data of > 450,000 users in Tokyo between across a 7-month period. 
Experiments show that the HR-SCI is capable of predicting COVID-19 outbreak hotspots 1-2 weeks before the official observations of the outbreak.
The HR-SCI metric can be used to develop early warning systems for COVID-19 hotspots that inform government agencies the locations of potential outbreaks, allowing them to plan effective prevention and preparation strategies.

The key contributions of this paper are as follows:
\begin{itemize}
    \item This study is the first to test the usage of web search and mobility data, which are linked by user IDs, for the prediction of COVID-19 outbreak hotspots.
    \item We propose a novel metric, \textbf{high risk social contact index (HR-SCI)}, that captures both the social contact density and the COVID-19 contraction risk levels of the users, with high spatio-temporal granularity.
    \item We verify that the \textbf{HR-SCI} can improve the predictability of COVID-19 hotspot locations through several case studies, compared to using just the social contact index or web search queries alone.
\end{itemize}

\section{Context: COVID-19 in Tokyo, Japan}
Japan has experienced a low number of cases and deaths due to COVID-19 in comparison to other countries in Europe and America, despite the social and physical proximity to China and intervention policies that are not as aggressive as some of the other countries \cite{dong2020interactive}. 
Non-pharmaceutical interventions implemented by the Japanese government include a non-mandatory closure and remote-working of non-essential business employees (February 26th), closures of public elementary, junior high and high schools (March 2nd), and inbound entry restrictions, starting with Hubei Province, China (February 3rd), until restricting inbound visitors from 73 countries (April 3rd). 
No mandatory lockdowns were enforced in Japan. 
The State of Emergency (SoE) was declared on April 7th, and was lifted on May 25th, after observing a significant decrease in cases (see Figure \ref{data_timeseries}, gray bars). 
Although Japan was able to contain the disease, many cities started to see an increase of cases in early July, and experienced the second wave in July to September (see Figure \ref{data_timeseries}, gray bars). 
Tokyo, which has the largest number of COVID-19 cases among prefectures in Japan, had around 400 new daily cases at its peak. 
As of October 17th, Tokyo has had 28,839 cases (out of 90,979 in Japan) and 434 deaths (out of 1,650 in Japan).  

\section{Data}
In this study, we utilized web search data and GPS location data, that are linked with the same user identifiers (IDs), owned by Yahoo Japan Corporation\footnote{https://about.yahoo.co.jp/info/en/}. 
Because the user IDs are linked, we are able to 1) identify users who have a high risk of COVID-19 contraction from their web search behavior, and 2) track their mobility patterns to measure their social contact rates with other users in the city. 

\subsection{Privacy Policies for User Data}
Yahoo Japan Corporation (YJ) has developed its own privacy policy and requires users to read and agree to its privacy policy before using any of the services provided by YJ.
Furthermore, because location data is highly sensitive for the users, users were asked to sign an additional consent form specific to the collection and usage of location data when they used apps that collect location information. 
The additional consent explains the frequency and accuracy of location information collection, and also the purpose and how the data will be used.
In addition to the above consent, YJ asked for additional consent from the users in this study because the analysis related to COVID-19 and personal health are much more sensitive.  
Therefore, YJ performed a double consent process, where the users who have given their consent to the usage of location information and web search queries were asked again, if they wish to provide their consent to be included in the dataset.

Moreover, YJ implemented strict restrictions in the analysis procedure. 
The methodology for handling the data and for obtaining user consent for this study were supervised by an advisory board composed of external experts.
YJ also ensured that research institutions other than YJ that participate in this study do not have direct access to the data. 
Although external research institutions were allowed to analyze aggregated data, the actual raw data were kept within YJ, and any analysis performed on raw data were performed within servers administered by YJ. 
In summary, given the high sensitivity of the study, this study was performed with careful consideration of the users' privacy.

\subsection{Individual Human Mobility Data}\label{mobdata}
GPS location data are anonymized so that individuals cannot be specified, and personal information such as gender, age and occupation are unknown. 
Each GPS location record contains the user's unique ID, timestamp of the observation, longitude, and latitude. 
The data has a sample rate of approximately 2\% of the entire population.
The data acquisition frequency of GPS locations varies according to the movement speed of the user to minimize the burden on the user's smartphone battery. 
If it is determined that the user is staying in a certain place for a long time, data is acquired at a relatively low frequency, and if it is determined that the user is moving, the data is acquired more frequently. 
We overcome this varying data acquisition frequency by spatially and temporally interpolating the location data, as explained in Section \ref{sci_method}.
A panel of users who were active each day in Tokyo metropolitan area before, during and after the COVID-19 pandemic were selected from the pool of users that have agreed to the aforementioned consents. 
This led to a sample of about 450,000 users, with approximately 50 observation points per user each day.

\subsection{Individual Web Search Data}
In addition to GPS location data, Yahoo Japan collects the web search queries of the users to improve the web search quality. 
Each web search query record contains the user's unique ID, timestamp of the search, and query text that was searched. 
The web search data were used to identify users who have a high risk of COVID-19 contraction, following the methods explained in Section \ref{scoring}.

\subsection{Number of COVID-19 Cases in Tokyo}
\begin{figure*}[t]
\centering
\includegraphics[width=\textwidth]{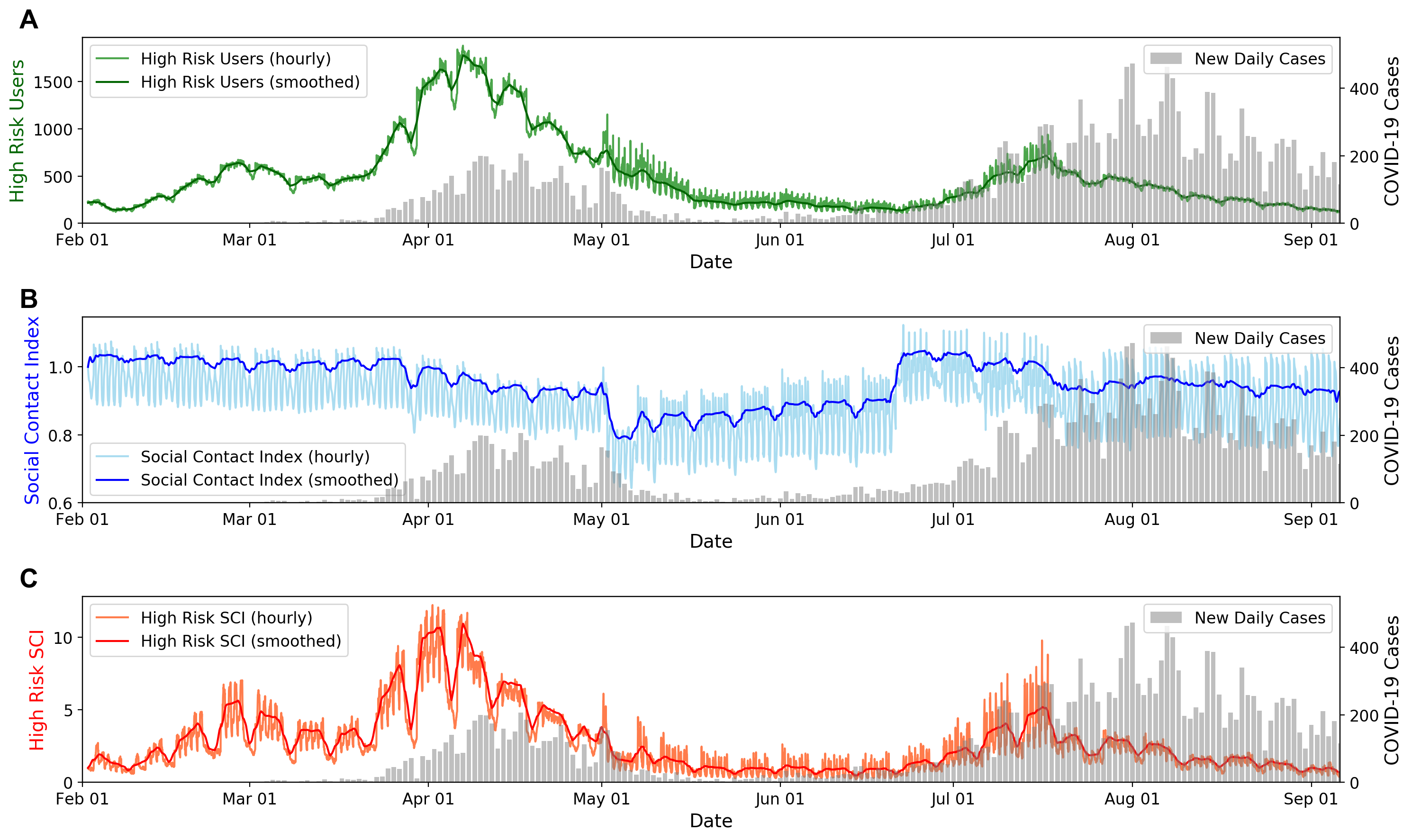}
\caption{Time series of estimated COVID-19 related metrics aggregated at the entire Tokyo metropolitan scale between February 1st and September 5th, 2020 (left axis). \textbf{A)} Total number of high risk users. \textbf{B)} Social contact index of an average user. \textbf{C)} High risk social contact index, which is the total amount of social contacts the high risk users encounter. Gray bars in each panel show the daily number of new COVID-19 cases in Tokyo (right axis).} 
\label{data_timeseries}
\end{figure*}

The daily number of confirmed COVID-19 cases in Tokyo were reported by the Tokyo metropolitan government through their Github data portal \cite{tokyometro} (gray bars in Figure \ref{data_timeseries}).

\section{Methodology}

\subsection{Preliminaries}
\subsubsection*{Definition 1 (\textbf{Web Search Session})} 
Each user's web search behavior can be observed as a sequence of web search queries performed by the user.
Usually, such continuous sequence of searches within a short time period are performed under a consistent underlying search intent. 
We define these short sequences of searches, which are assumed to be under a consistent search intent, as ``web search sessions''.
Web search sessions are used to detect high risk users who intensively search about COVID-19 symptoms (Section \ref{scoring}).

\subsubsection*{Definition 2 (\textbf{High Risk User})} 
A high risk user is defined as a user who conducts more web search sessions that are related to COVID-19 symptoms than the pre-defined threshold. 

\subsubsection*{Definition 3 (\textbf{Social Contact Index})} 
The social contact index (SCI) measures the average amount of encounters that each user experiences due to movements outside of their estimated home locations. 

\subsubsection*{Definition 4 (\textbf{High Risk Social Contact Index})} 
The high risk social contact index (HR-SCI) is a composite measure of the SCI and the number of high risk users. 
HR-SCI measures the average amount of social contacts that high risk users encounter due to movements outside of their estimated home locations. 

\subsection{Scoring of Web Search Queries}\label{scoring}
To identify high risk users of COVID-19 contraction, the users' web search queries from Yahoo Japan Search were given risk scores.
Queries submitted by users were treated as ``Covid-19 queries'' if they match against pre-defined query patterns.
The query patterns consist of 3 types of query phrases: 1) COVID-19 symptom related queries, 2) names of medical institutions related to COVID-19 care, and 3) names of locations, as shown in Table \ref{queries}. 
For the first group of queries, 186 COVID-19 symptom related queries were determined, such as 'coronavirus high fever' or 'may have coronavirus', including various slang words (e.g., 'corona' instead of 'coronavirus', which is typically used by Japanese-speaking users). 
For the second group of queries, 5 queries that represent medical institutions concerned with COVID-19 were determined. 
These include facilities such as hospitals designated by the local health authorities to be specialized for the treatment of COVID-19 infected patients.
The third group of queries include 2168 queries representing names of locations (e.g., Shibuya).
The latter two groups of queries only were scored as ``COVID-19 related'' only when they were searched together (e.g., ``Central hospital Shibuya'').
To ensure the fairness of future analyses and to avoid users from making these searches even when they do not have COVID-19 symptoms, we do not disclose the details of the query list.
When a user has at least one COVID-19 related search query within a session, we determine that web search session as a COVID-19 related web search session.
Given a pre-defined threshold $k$, a user is identified as a high risk user is he/she had more than $k$ COVID-19 related web search sessions.  
Figure \ref{data_timeseries}A shows the hourly number of high risk users with $k=3$ in the Tokyo region, along with the daily number of new cases.

\subsection{Social Contact Analysis}\label{sci_method}
Using the GPS location data of each user, we compute the social contact index within a given region on the time  of interest. 
To overcome data sparsity caused by battery saving functions of the app (explained in Section \ref{mobdata}), we perform spatio-temporal interpolation of the GPS location observations. 
Because the GPS data are collected less frequently when movement is detected, we assume that the individual users are static while there are no observations. 
Using the interpolated individual trajectory data produced from mobile phone location data, the social contact indexes were computed. 
The social contact indexes shown in Figure \ref{data_timeseries}B were computed for 30 minute intervals. 
First, for each time interval $[t,t+dt)$, where $dt=30$ minutes, users who were not within 125 meters from their estimated home locations (via meanshift clustering of nighttime staypoints) were detected as ``staying out''.
We denote this set of individual users as $N^{\rm out}_t$. 
For user $i$ staying out ($i \in N^{\rm out}_t$), we compute the number of other users ``staying out'' who are within 125 meters from user $i$, and use that count $c_{i,t}$ as a proxy for social contacts. 
The social contact index is calculated as the total social contacts for all users staying out, divided by the total number of users including those staying at their homes. 
Thus, mean social contact value is computed as $C_t = 
\sum_{i\in N^{\rm out}_t} c_{i,t}/N$, where $N$ is the total number of users observed on that day. 
The social contact index is the relative value of mean social contacts with respect to typical mobility patterns, observed before the COVID-19 pandemic. 
Thus, the social contact index (SCI) of 1 corresponds to the same amount of social contacts as the daily peak times on weekdays before the COVID-19 pandemic. 

Figure \ref{data_timeseries}B shows the SCI between February 1st and September 5th in metropolitan Tokyo. 
We observe a gradual decrease starting from April, low SCI during May and June, and a rise in SCI in July and August. 
By comparing with the new daily cases (gray bars), SCI reduces significantly after the first wave but the decrease is more subtle after the second wave. 
Figure \ref{data_timeseries}C shows the high risk SCI (HR-SCI). 
We observe that similar to the number of high risk users (Figure \ref{data_timeseries}A), the number of high risk users increase slightly before both the first and second waves, however, the increase is smaller during the second wave. 
In the following sections, we investigate the effectiveness of these three metrics in predicting COVID-19 outbreaks in different geographical scales (macroscopic and microscopic). 

\begin{figure*}[t]
\centering
\includegraphics[width=\textwidth]{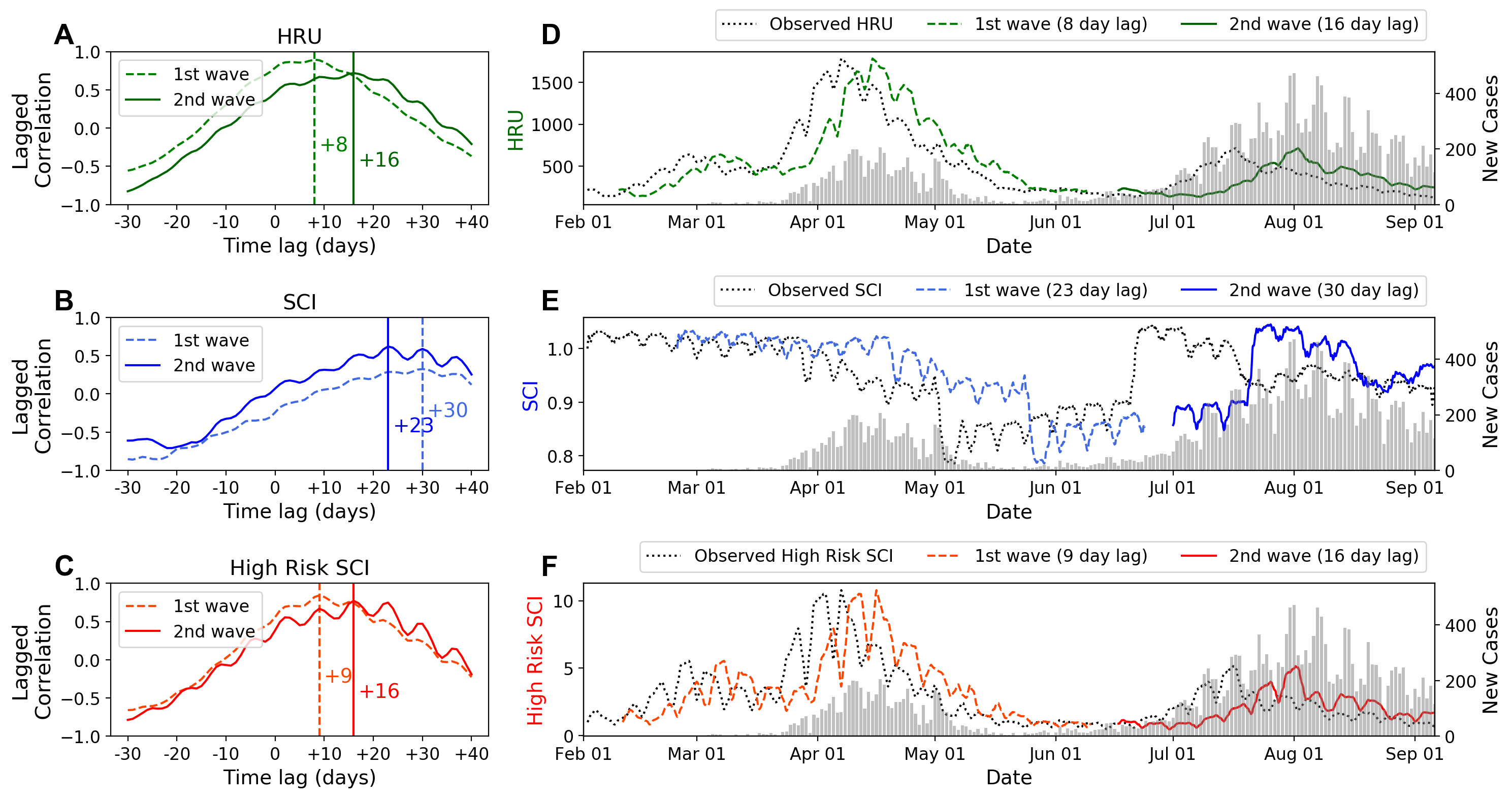}
\caption{Time lagged cross correlation analysis of the high risk users, social contact index, and high risk social contact index metrics against the daily number of new cases. Results indicate higher predictability using HRU and HR-SCI; the metrics preceded the daily cases trend by 8-9 days during the first wave, and by 16 days during the second wave.} 
\label{corr_lag}
\end{figure*}

\begin{table}
\small\sf\centering
\caption{Query words that were used for scoring users' risk level using web search data.}
\begin{tabular}{lll}
\toprule
Type & Quantity & \begin{tabular}{l}Examples\end{tabular} \\
\midrule
symptom related  & 186 & \begin{tabular}{l}``coronavirus high fever'' \\ ``no smell'', ``may have corona'' \end{tabular} \\
\midrule
medical institutions & 5 & \begin{tabular}{l}``hospital'', ``medical clinic''\end{tabular}
\\
\midrule
location names & 2168 & \begin{tabular}{l}``Shibuya'', ``Shinjuku''\end{tabular}
\\
\bottomrule
\multicolumn{3}{l}{\begin{tabular}{l}*queries on medical institution names and location names need to\\ be searched together to qualify as a ``COVID-19 related search''\end{tabular}}
\end{tabular}
\label{queries}
\end{table}

\section{Results}
In this section, we investigate the effectiveness of each metric (number of high risk users, social contact index, high risk social contact index) in predicting COVID-19 outbreak hotspot locations on different spatial scales. 
Sections \ref{macro} and \ref{micro} investigate the predictability of the macroscopic (entire Tokyo metropolitan area) and microscopic (125 meters $\times$ 125 meters level) trends of new COVID-19 cases, respectively.

\begin{table}
\small\sf\centering
\caption{Time lagged cross correlation and lagged days (in brackets) of the three metrics with the total daily number of COVID-19 cases in Tokyo.}
\begin{tabular}{clll}
\toprule
 & \multicolumn{3}{c}{Pearson Correlation (lagged days)} \\
 \cmidrule{2-4}
 & High Risk Users & SCI & High Risk SCI  \\
\midrule
1st wave & \textbf{0.894} (+8 days)  & 0.324 (+30 days) & 0.840 (+9 days) \\
2nd wave & 0.719 (+16 days) & 0.619 (+23 days) & \textbf{0.770} (+16 days) \\
\bottomrule
\end{tabular}
\label{macrotable}
\end{table}

\subsection{Macroscopic Trend Prediction}\label{macro}
From visual inspection of Figure \ref{data_timeseries}, we observe that the trends of high risk users (panel A) and high risk social contact index (panel C) have two peaks in early April and mid July, similar to the trends of daily new COVID-19 cases, which also has two peaks.  
To quantify the predictability of daily cases trend, we computed the time lagged cross correlation between the metric time series data and the daily cases trend. 
Since we observe different patterns during the first and second waves, we divide the daily cases trend into the first wave (February 1st $\sim$ May 31st) and the second wave (June 1st $\sim$ September 5th), and compute the lag for each of the periods. 
A positive lag would indicate that the metric precedes the daily number of cases, while a negative lag would indicate the opposite. 

Figure \ref{corr_lag}A-C shows the estimation results of the temporal lag for the three metrics: high risk users (HRU), social contact index (SCI), and high risk social contact index (High Risk SCI). 
The results indicate that the two metrics -- high risk users and high risk social contact index -- have similar high predictability of the daily number of cases. 
The time lagged time series data are visualized in Figure \ref{corr_lag}D-F, where the dotted black line shows the original observation, and colored dashed and solid lines show the time lagged first wave and second wave predictions, respectively. 
The peak cross correlation are shown in Table \ref{macrotable}, and indicates that while HRU and HR-SCI metric perform better, SCI itself is not able to capture the trends of the number of new cases. 
The HRU metric performed the best during the first wave with $R=0.894$ and lag of 8 days, while HR-SCI performed the best during the second wave with $R=0.770$ and a lag of 16 days.
From these results, we conclude that both the HRU and HR-SCI metrics perform well to predict the temporal trends of new daily COVID-19 cases.

\begin{figure*}
\centering
\includegraphics[width=0.7\columnwidth]{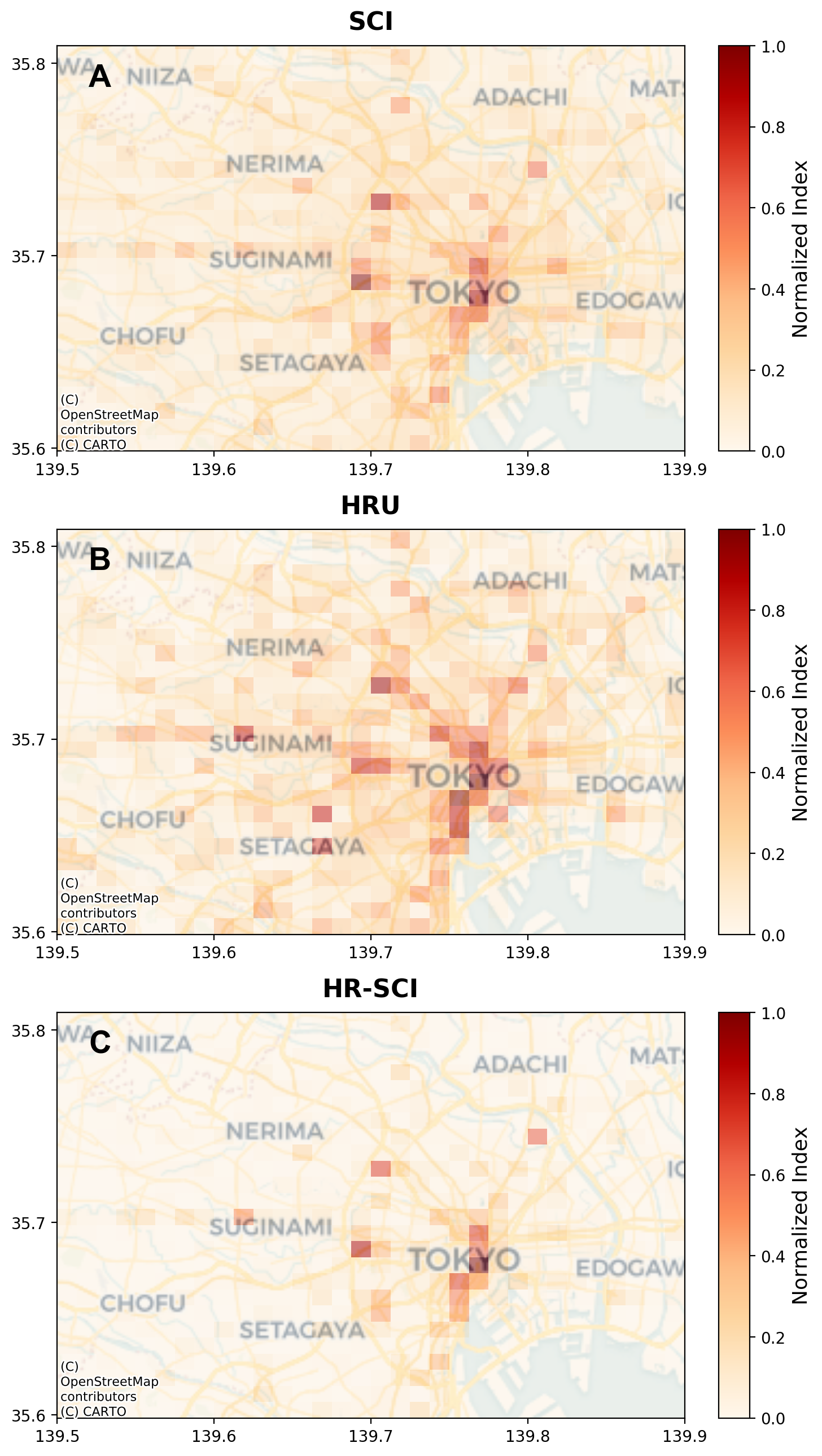}
\caption{Spatial plot of the total (A) social contact index, (B) high risk user count, and (C) high risk social contact index during June 1st - July 31st in the metropolitan Tokyo area, aggregated into 1km grid cells.} 
\label{mapfig}
\end{figure*}

\begin{table}
\small\sf\centering
\caption{Top 10 locations in Tokyo metropolitan area with highest SCI, HRU, and HR-SCI indexes. }
\begin{tabular}{clll}
\toprule
Rank & SCI & High Risk Users & High Risk SCI  \\
\midrule
1 & West Shinjuku & Hibiya & Tokyo \\
2 & Tokyo & Tokyo & West Shinjuku \\
3 & Ikebukuro & Ikebukuro & Akihabara \\
4 & Akihabara & Akihabara & Hibiya \\
5 & Ginza & Hamamatsu-cho & Ikebukuro \\
6 & Ohtemachi & Shimbashi & Kitasenju \\
7 & Hibiya & Shimokitazawa & Ohtemachi \\
8 & Kitasenju & West Shinjuku & Shimbashi \\
9 & South Shibuya & Ogikubo & Ogikubo \\
10 & Shinagawa & Iidabashi & Hamamatsu-cho \\
\bottomrule
\end{tabular}
\label{top10}
\end{table}

\subsection{Microscopic Outbreak Hotspot Prediction}\label{micro}
In the previous section, it was shown that HR-SCI and HRU indexes were both effective in predicting the macroscopic (Tokyo metropolitan area scale) prediction of COVID-19 outbreak. 
However, policy makers could benefit from more microscopic, finer spatially-grained prediction (and early warning system) of outbreak hotspots. 
In this section, we tested the effectiveness of the three metrics on predicting outbreak hotspots beforehand, in the microscopic (i.e., 125 meters grid) spatial scale.  

\begin{figure}[t]
\centering
\includegraphics[width=.6\columnwidth]{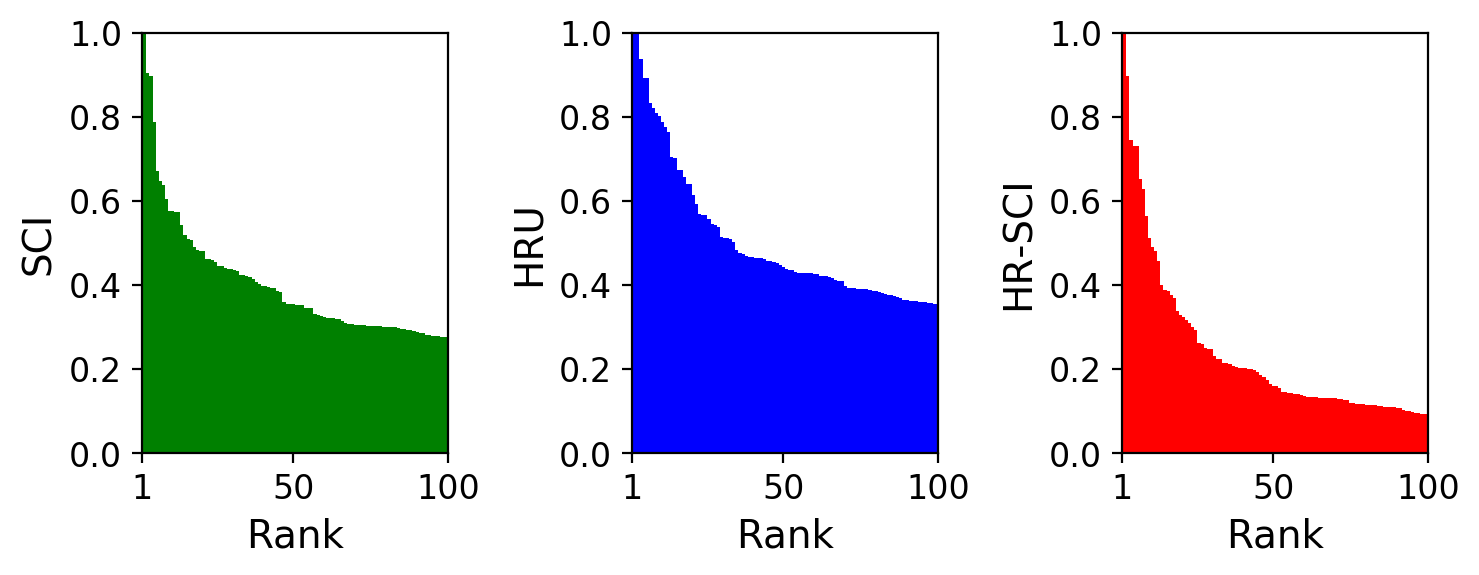}
\caption{Normalized scores of each metric in 1km grid cells sorted by their rank, showing high skewness of the HR-SCI.} 
\label{skew}
\end{figure}

\begin{figure*}[t]
\centering
\includegraphics[width=\textwidth]{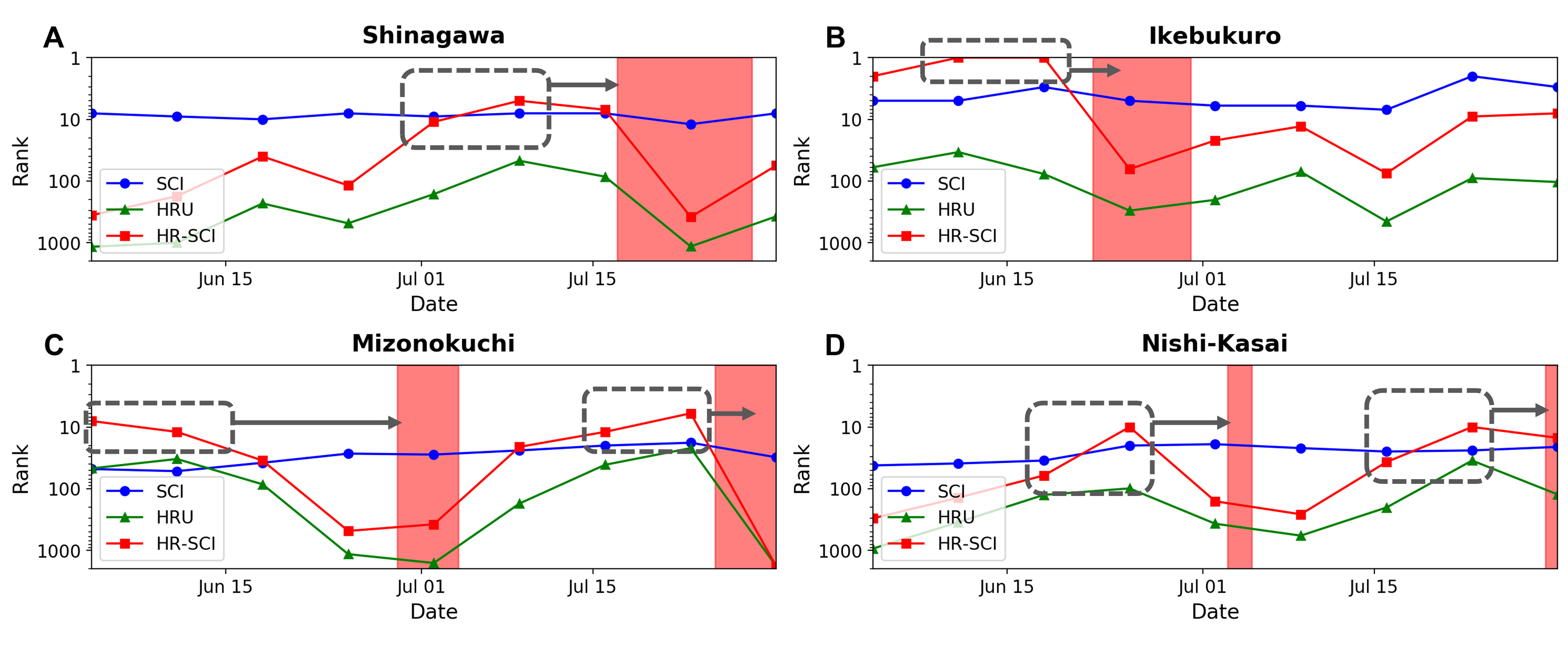}
\caption{The rank of four areas in Tokyo using social contact index, high risk users index, and high risk social contact index between early June and end of July. Red shaded periods show local COVID-19 outbreak timings. Arrows show that HR-SCI is able to forecast outbreaks 1-2 weeks beforehand.} 
\label{micro_casestudy}
\end{figure*}

First, we estimate the three indexes (SCI, HRU, and HR-SCI) for each 1km $\times$ 1km grid in Tokyo metropolitan region using similar procedures as explained in the previous section. 
Figure \ref{mapfig} shows the spatial distributions of the three index scores (SCI, HRU, HR-SCI) on the 1km $\times$ 1km grid scale, aggregated between June 1st and July 31st, and normalized between 0 and 1.
From Table \ref{top10}, we observe high SCI scores in major central business districts including Shinjuku, Tokyo, and Ikebukuro, whereas the high risk index is more dispersed, with lower density areas such as Hibiya, Hamamatsu-cho, and Ogikubo ranked higher compared to SCI. 
HR-SCI is a composite measure, including locations from both the SCI and HRU ranking. 
As we can see from the normalized scores of the top 100 ranked locations of each metric in Figure \ref{skew}, HR-SCI is more selective, showing high scores on only a few grid cells.
For the grid cells that ranked in the top 50 out of the $\sim$ 2000 cells in any of the three metrics, we further compute the metrics on  the 125m $\times$ 125m scale to obtain a more precise analysis on outbreak hotspot predictability. 

To validate the effectiveness of the HR-SCI scores on predicting outbreak hotspot locations, we collected news articles reporting the occurrence of COVID-19 outbreaks in Tokyo. 
We found news reports of outbreak events in four areas in Tokyo -- Shinagawa, Mizonokuchi, Ikebukuro, and Nishi-Kasai -- which were used to validate the predictability of outbreak hotspots using the metrics (see Table \ref{reports} for information sources).

\subsubsection*{\textbf{Shinagawa area}} 
Shinagawa (Figure \ref{micro_casestudy}A) is known as one of the largest central business districts in Tokyo, and therefore consistently has a high social contact index, ranking between 8th to 10th in all periods. 
However, we can also observe that Shinagawa ranks very low (between 100th to 1000th) using the high risk user count index. 
On the other hand, Shinagawa's high risk social contact index rank fluctuates over time, and ranks 5th (week of July 6th) and 7th (week of July 13th). 
It is reported that cases in Shinagawa increased from July 13th to August 2nd, and has had several COVID-19 clusters during that period.  
Therefore, using SCI would result in false positive predictions until mid July, whereas HR-SCI effectively predicts outbreaks with 2 weeks in hand.

\subsubsection*{\textbf{Ikebukuro area}} 
Similar to the Shinagawa area, Ikebukuro is also a bustling area with many shopping and business facilities, and ranks constantly in the top 10 using SCI (between 2nd and 7th in entire period) (Figure \ref{micro_casestudy}B).
However, Ikebukuro never ranks in the top 50 using high risk users index. 
Using HR-SCI, Ikebukuro ranks 2nd, 1st, 1st in the three weeks from June 1st, and again ranks in the top 10 during the end of July (weeks of July 20th and 27th). 
It has been reported that in Ikebukuro, the number of cases have risen from mid-June to the end of June, and by the end of June had the most number of new cases in Tokyo.
Moreover, occurrence of COVID-19 clusters in large shopping malls and the city hall have been reported at around early August (which is 1-2 weeks after the second increase in HR-SCI).
Ikebukuro showcases another example where using SCI, it would constantly be false-positively predicted as outbreak locations, but using HR-SCI, outbreak timings can be accurately predicted 1-2 weeks beforehand.

\subsubsection*{\textbf{Mizonokuchi area}} 
Mizonokuchi area is a residential area located in the suburbs of Tokyo, unlike Shinagawa and Ikebukuro. 
As shown in Figure \ref{micro_casestudy}C, Mizonokuchi never ranks in the top 10 using social contact index, due to low population density. 
However, we observe an increase in HR-SCI in early June and the week of July 20th (ranked 6th in the entire Tokyo region), which coincides with the two outbreaks of COVID-19 in the area reported at early July and from the end of July to beginning of August. 
The case study of Mizonokuchi shows that even in low population density areas, HR-SCI is able to predict outbreaks with 2 weeks  beforehand. 

\subsubsection*{\textbf{Nishi-Kasai area}} 
Nishi-Kasai area, similar to Mizonokuchi area, is a residential area with smaller active population compared to Shinagawa and Ikebukuro. 
Therefore, the area always ranks below 20th in SCI. 
However, using the HR-SCI, Nishi-Kasai ranks in the top 10 twice during the study period: the weeks of June 22nd and July 20th (Figure \ref{micro_casestudy}D). 
Indeed, COVID-19 outbreaks have been reported on July 4th and July 30th in the area, which are both 2 weeks after the area was ranked 10th using the HR-SCI. 
Nishi-Kasai is another case where because of the low active population density, the area never appears in the top rankings using SCI, but accurately appears before actual outbreaks using HR-SCI. 

\begin{table*}[h]
\small\sf\centering
\caption{News reports of local COVID-19 outbreaks in the four locations.}
\begin{tabular}{m{2cm} m{13cm}}
\toprule
Location & Information Source \\
\midrule
Shinagawa & 
\begin{minipage}[c]{150.0mm}
\vspace{.2cm}
\begin{itemize}
    \item COVID-19 cluster at a pub in Shinagawa on July 17th - July 28th, 2020
    \begin{itemize}
        \item \url{https://www.city.shinagawa.tokyo.jp/PC/press_release/press_release-2020/20200727192108.html}
        \item \url{https://www.yomiuri.co.jp/national/20200728-OYT1T50257/}
    \end{itemize} 
\end{itemize} \end{minipage} \vspace{.2cm} \\
\midrule
Ikebukuro & 
\begin{minipage}[c]{150.0mm}
\vspace{.2cm}
\begin{itemize}
    \item News of "The number of infected people in Ikebukuro increased at the end of June".
    \begin{itemize}
        \item \url{https://news.yahoo.co.jp/articles/ab1015d5df8491b18e65cd4a9d8c3a9ed16c8170}
        \item \url{https://www.nikkei.com/article/DGXMZO60974500Q0A630C2CC1000/}
        \item \url{https://www3.nhk.or.jp/news/html/20200703/k10012494661000.html}
        \item \url{https://www.tokyo-np.co.jp/article/39361}
    \end{itemize} 
\end{itemize}  \end{minipage} \vspace{.2cm} \\
\midrule
Mizonokuchi & 
\begin{minipage}[c]{150.0mm}
\vspace{.2cm}
\begin{itemize}
    \item Rapid increase in the number of cases in late June and early July
    \begin{itemize}
        \item  \url{https://www.city.kawasaki.jp/350/page/0000115886.html}
        \item  \url{https://hiyosi.net/2020/07/03/corona_virus-46/}
    \end{itemize} 
    \item High number of corona cases in early August
    \begin{itemize}
        \item \url{https://news.yahoo.co.jp/articles/152b778e3630b9859a8ded10476e388eef721156}
        \item \url{https://hiyosi.net/2020/07/27/corona_virus-51/}
        \item \url{https://www.pref.kanagawa.jp/docs/y4g/prs/r3008440.html}
        \item \url{High number of corona cases in early August}
    \end{itemize} 
\end{itemize}   \end{minipage} \vspace{.2cm} \\
\midrule
Nishi-Kasai & 
\begin{minipage}[c]{150.0mm}
\vspace{.2cm}
\begin{itemize}
    \item COVID-19 cluster at a office in Nishi-Kasai on July 4th, 2020
    \begin{itemize}
        \item \url{https://www.city.edogawa.tokyo.jp/e004/bosaianzen/covid-19/ncov-jigyou/20200714.html}
    \end{itemize} 
    \item COVID-19 cluster at a pub in Nishi-Kasai on July 30th, 2020
    \begin{itemize}
        \item \url{https://www.news24.jp/articles/2020/08/12/07698857.html}
        \item \url{https://www.city.edogawa.tokyo.jp/e004/bosaianzen/covid-19/ncov-jigyou/20200812.html}
    \end{itemize} 
\end{itemize}   \end{minipage} \vspace{.2cm} \\
\bottomrule
\multicolumn{2}{l}{*All webpages were accessed on October 18th, 2020}
\end{tabular}
\label{reports}
\end{table*}

\section{Discussion}

\subsection{Pros and Cons of Proposed Metrics}
In this study, we investigated the effectiveness of various metrics for predicting outbreak hotspot locations of COVID-19, using a unique dataset that which contains both the web search queries and GPS location information of users. 
The two experiment setups -- predicting the macroscopic (Tokyo metropolitan area) and microscopic (oubreaks in 125 meter cell scale) -- unraveled the strengths and weaknesses of the three metrics: high risk user (HRU) count, social contact index (SCI), and high risk social contact index (HR-SCI). 

It was shown in Figure \ref{corr_lag} and Table \ref{macrotable} that HRU is effective in capturing the macroscopic trends of COVID-19 outbreaks. 
We were able to observe high correlation between the daily number of mobile phone users who had high number of COVID-19 related web search sessions and the daily new cases count provided by the Tokyo Metropolitan Government. 
In particular, cross correlation analysis showed that the fluctuations in the high risk user count preceded the cases count by 1-2 weeks, which can be useful in predicting the outbreaks in the macroscopic scale. 
However, HRU index was not effective in predicting outbreak locations in the microscopic scale, as shown in Section \ref{micro}. 
This is because on the microscopic scale (125 meter grid cells), the decrease in the average number of users observed in each cell increases the noise in the data. 
These metrics could become biased even by small groups of people who search about COVID-19 out of pure interest. 

The social contact index (SCI), which is used in many studies, could be effective in predicting outbreaks if COVID-19 cases occur in high dense areas, since essentially SCI captures the population density of that area.
However, as shown in the analysis in Section \ref{micro}, SCI has two major drawbacks in predicting hotspots. 
First, the SCI is relatively static over time, staying consistently large in locations that are more congested, including central business districts and hub stations (e.g., Tokyo, Shibuya, Shinjuku), and consistently low in more rural areas. 
Although it is true that COVID-19 has a higher probability of spreading in high-density areas, because of this static nature of the SCI metric, it is difficult to predict the timings of such outbreaks. 
In the case of Shinagawa (one of the busiest central business districts in Tokyo), outbreaks could not be detected using SCI because the area was always congested throughout the COVID-19 crisis. 
Mizonokuchi (an town in the outskirts of Tokyo) which had an outbreak also could not be detected using SCI because of the consistently low SCI. 

Considering the pros and cons of the HRU and SCI metrics, the high risk SCI (HR-SCI) metric is able to capture both 1) existence of high risk users and 2) high population (contact) density.
Thus, especially in the microscopic scale, HR-SCI was shown to be effective in predicting hotspot locations in Section \ref{micro}. 
Therefore, HR-SCI could be utilized to implement early warning systems for COVID-19 outbreak hotspot locations.

\subsection{Limitations and Future Works}
The presented empirical results should be considered in the light of some limitations.
First, we found that although the HRU and HR-SCI metrics had high predictability of COVID-19 cases in Figure \ref{corr_lag}, the optimal time lags in the two waves varied (first wave: 9 days, second wave: 16 days). 
Many reasons could be causing this difference, however, this suggests a change in the the web search behavior of the users between the two waves. 
As the virus spreads and more information becomes available due to the increase of cases, the type of symptoms that users search in relation to COVID-19 could increase (e.g., it was unknown that loss of smell was related to COVID-19 in the first weeks of the outbreak). 
Moreover, the difference in the absolute amount of high risk users between the two waves show that awareness levels of COVID-19 had also changed, and that as people become more used to the disease, less people search about COVID-19 symptoms. 
Despite these issues, the finding that we can predict the outbreak around 1-2 weeks beforehand is valuable and has various potential applications as discussed later. 
Future research could focus more on the modeling of the web search behavior to improve the accuracy of the outbreaks. 

Second, computational costs increase drastically as we increase the spatial resolution of the analysis. 
For example, in Tokyo, there are around 2400 1km $\times$ 1km grid cells, which were further divided into over 150,000 125m $\times$ 125m grid cells. 
To cope with this issue, we first ran the analysis on the 1km grid scale, and further conducted analysis on 125m grid cells only in selected high ranked cells. 
However, we could be missing microscopic locations with high risk scores if they are surrounded with low scores. 
Running the analysis with high computing resources could be one solution to this issue. 

Finally, collecting ground truth data for COVID-19 outbreaks on the microscopic spatial scales were challenging, with information sources scattered across websites of various institutions and agencies. 
Constructing a comprehensive database that contains when, where, and how many patients were reported on fine spatial scales (with careful consideration on privacy concerns) would be valuable for future research. 

\subsection{Potential Applications}
The usability of high risk social contact index is not limited to COVID-19, and can be applied to any contagious disease in theory. 
For example, flu has been the subject of many previous studies on predicting disease spread using web search data and mobility data (e.g., \cite{lampos2010tracking,panigutti2017assessing}). 
Although the web search intensity, frequency, and the disease related parameters (e.g., effective reproduction number) could be different in other diseases, it would be valuable to test this method on other epidemics.
Moreover, testing the applicability of this method in different regions around the world could be an interesting extension to this study.

In addition to the early warning system mentioned previously, these metrics can be used to develop a personalized alert and navigation app on smartphones. 
Although there are existing apps that inform users about their risk of contracting the disease based on their mobility information (e.g., COCOA app\footnote{\url{https://www.mhlw.go.jp/stf/seisakunitsuite/bunya/cocoa_00138.html}} developed by the Japanese Government), they do not provide early warning information on where potential outbreaks could occur. 
Using the high risk social contact index, we would be able to provide navigation and personalized early warning to users so that they can avoid visiting or passing through high risk areas.

\section{Related Works}
\subsection{Mobility Analysis during COVID-19}
Mobile devices have become ubiquitous in many areas around the world, providing opportunity to analyze human mobility dynamics in an unprecedented spatial and temporal granularity and scale \cite{blondel2015survey}.
To monitor and evaluate the impacts of such interventions, large scale mobile phone data has been utilized as an effective data source \cite{oliver2020mobile}. 
Studies on human mobility analysis have used such data to model disease dynamics \cite{bengtsson2015using,finger2016mobile,tizzoni2014use,wesolowski2012quantifying}.
During the COVID-19 crisis, researchers, industry, and government agencies have utilized large-scale mobility datasets to evaluate the effects of NPIs in various countries, including the United States \cite{gao2020mapping,klein2020assessing,wellenius2020impacts}, the United Kingdom \cite{santana2020analysis}, Italy \cite{bonato2020mobile,pepe2020covid,cintia2020relationship,bonaccorsi2020economic}, China \cite{kraemer2020effect,lai2020effect}, Sweden \cite{dahlberg2020effects}, Germany \cite{schlosser2020covid19}, Spain \cite{orro2020impact}, Austria \cite{heiler2020country}, and Japan \cite{yabe2020non,mizuno}.
None of these studies have attempted to integrate the analysis of web search queries to predict COVID-19 outbreak hotspots. 

\subsection{Applications of Web Search Data Analysis}
Mining of web search query data has attracted the attention of researchers and practitioners ever since search engines were introduced to the world \cite{silverstein1999analysis}. 
Web search data has been utilized for various applications, for example, to predict users' demographics \cite{wu2019neural} and mobility decisions during crisis \cite{yabe2019predicting}. 
During the COVID-19 pandemic, many studies have utilized web search query data to understand information seeking behavior and the occurrence of infodemics \cite{rovetta2020covid,bento2020evidence,mavragani2020tracking}. 
Others have used such data to detect the increase in COVID-19 symptoms (e.g. loss of smell) \cite{walker2020use, rajan2020association}, and also for predicting outbreaks \cite{li2020retrospective,hisada2020syndromic}. 
With the availability of open datasets \cite{xu2020epidemiological}, there is great potential in further using web search query data for pandemic response and prevention. 
Despite the vast array of studies, none have attempted to integrate web search data with mobility data to predict COVID-19 outbreaks. 

\section{Conclusion}
As COVID-19 continues to affect public health in cities across the world, early warning systems that can predict where the next outbreak would occur is of significant importance for government agencies. 
In this study, we used web search data and GPS location data, which are linked with common user IDs, to predict outbreak hotspot locations using the \textit{high risk social contact index}. 
Validation using data from Tokyo, Japan showed that compared to previously proposed metrics, the high risk social contact index is capable of predicting the timing of outbreaks 1-2 weeks beforehand in a microscopic (125 meters) spatial scale.
This study proposes a novel method to predict disease outbreak hotspots, which can be used to develop early warning systems that may assist government agencies to prepare effective strategies for disease spread prevention.

\section*{Acknowledgements}
The authors would like to thank Zengxiang Lei, PhD student in the Lyles School of Civil Engineering at Purdue University for assisting data collection.

\bibliographystyle{unsrt}  
\bibliography{references}  


\end{document}